\documentclass{WileyMSP-template}
\usepackage{amsmath}
\usepackage{mathrsfs}
\begin{document}

\title{Deterministic and universal quantum squeezing gate with a \\teleportation-like protocol}

\maketitle

\author{Xiaocong Sun}
\author{Yajun Wang}
\author{Yuhang Tian}
\author{Qingwei Wang}
\author{Long Tian}
\author{Yaohui Zheng*}
\author{Kunchi Peng}

\begin{affiliations}
Xiaocong Sun, Yajun Wang, Yuhang Tian, Qingwei Wang, Long Tian, Prof. Yaohui Zheng, Prof. Kunchi Peng\\
State Key Laboratory of Quantum Optics and Quantum Optics Devices, Institute of Opto-Electronics, Shanxi University, Taiyuan 030006, China\\
Email Address: yhzheng@sxu.edu.cn

Yajun Wang, Long Tian, Prof. Yaohui Zheng, Prof. Kunchi Peng\\
Collaborative Innovation Center of Extreme Optics, Shanxi University, Taiyuan 030006, China

\end{affiliations}


\keywords{quantum informtion, quantum gate, quantum protocol}

\begin{abstract}

Squeezing transformation as an essential component, gives rise to the possibility to perform various tasks of quantum information processing. However, the reported squeezing gate with best performance so far is a conditional realization at the expense of low success probability, while the performance of the deterministic ones is currently circumscribed by the limited squeezing degree of the non-classical ancilla. To address this issue, we develop and demonstrate a new scheme of deterministic and universal quantum squeezing gate with the property of non-local operation attributed to the teleportation-like protocol. We demonstrate high fidelity squeezing operation, even when the level of target squeezing being up to 10 dB, where a squeezed state with non-classical noise reduction of 6.5 dB is directly observed. Moreover, we perform a high-fidelity complex operation including a Fourier transformation and a phase squeezing gate, exploring the potential of implementing the complex task of quantum information processing with the presented functional unit. Our method can be applied to distributed quantum processor, the creation of exotic non-classical states, and quantum error correction.

\end{abstract}

\section{Introduction}
Gaussian transformation along with Gaussian state preparation and tomography constitutes the continuous variable (CV) quantum information processing (QIP) \cite{RMPQICV,RMP,CVQIP}, wherein squeezing gate plays an essential role. Squeezing gate offers remarkable opportunities for realizing CV universal quantum computation \cite{Mile,MBQC,sxlQC,UkaiPRA,FurusawaSG,FurusawaQPG,FurusawaDSG}. In addition, high-fidelity squeezing transformation is desirable in quantum error correction with Gottesman-Kitaev-Preskill scheme \cite{GKP,GKP_SQ,GKP_QEC,GKP_FTQEC}. 

Although 15 dB-squeezed vacuum state in optical domain has been generated via optical parametric oscillator (OPO) \cite{Schnabel}, in-line squeezing operation with parametric process is not feasible to general input state, since this operation on the fragile quantum states is impeded by the decoherence due to the ineluctable practical imperfections. To pursue the goal of high performance squeezing gate, two off-line protocols were developed and proof-of-principle demonstrated by tailoring the auxiliary resource states. Among them, a recent realization of heralded squeezing gate exploiting an ancillary squeezed state \cite{PKL,LiuOE}, reported the best performance so far with respect to measurement-based feed-forward operation \cite{Furusawa_squeezing,SXL_EPR,Furusawa_cluster}, but at the expense of rather low success probabilities. Furthermore, higher the target squeezing, lower the success probabilities attributed to the narrower post-selection filter window (target squeezing of 10 dB with success probability of $10^{-4}$). In addition, this protocol presents a inherent property of local operation, thus obstacles the application in the distributed quantum processor, one approach to large-scale quantum computer \cite{Jiang,Eisert,Chou,Rempe}. 

Here, we present a scheme to attain an unconditional quantum squeezing gate with high-fidelity. Incorporating the high quality Einstein-Podolsky-Rosen (EPR) entangled resource, we demonstrate the ability to operate the squeezing gate with high fidelity even when the target squeezing is set up to 10 dB, much better than the record in previous deterministic configurations (target squeezing of 6.0 dB with a fidelity of 78\% reported in Ref. \cite{Furusawa_squeezing}). Strikingly, in contrast to conventional entanglement ancilla-assisted feed-forward configuration \cite{SXL_EPR,Furusawa_cluster}, only optical splitting ratio and electronic feedback gain are optimized corresponding to various target squeezing, relaxing the technical challenge in relative phase locking at arbitrary angles. Compared with the protocol with the ancillary squeezed state, the one presented here inherently promises non-local operation, offering the opportunities to implement the distributed quantum tasks. Furthermore, we take the complex operation by combining the squeezing gate with a Fourier gate as an example to illustrate the potential of implementing the complex task of quantum information processing with high fidelity by means of the presented scheme, which overcomes the challenge met by the scheme employing an ancillary squeezed state \cite{SI}.

\section{Quantum protocol}
\begin{figure}
  \centering
  \includegraphics[width=0.6\linewidth]{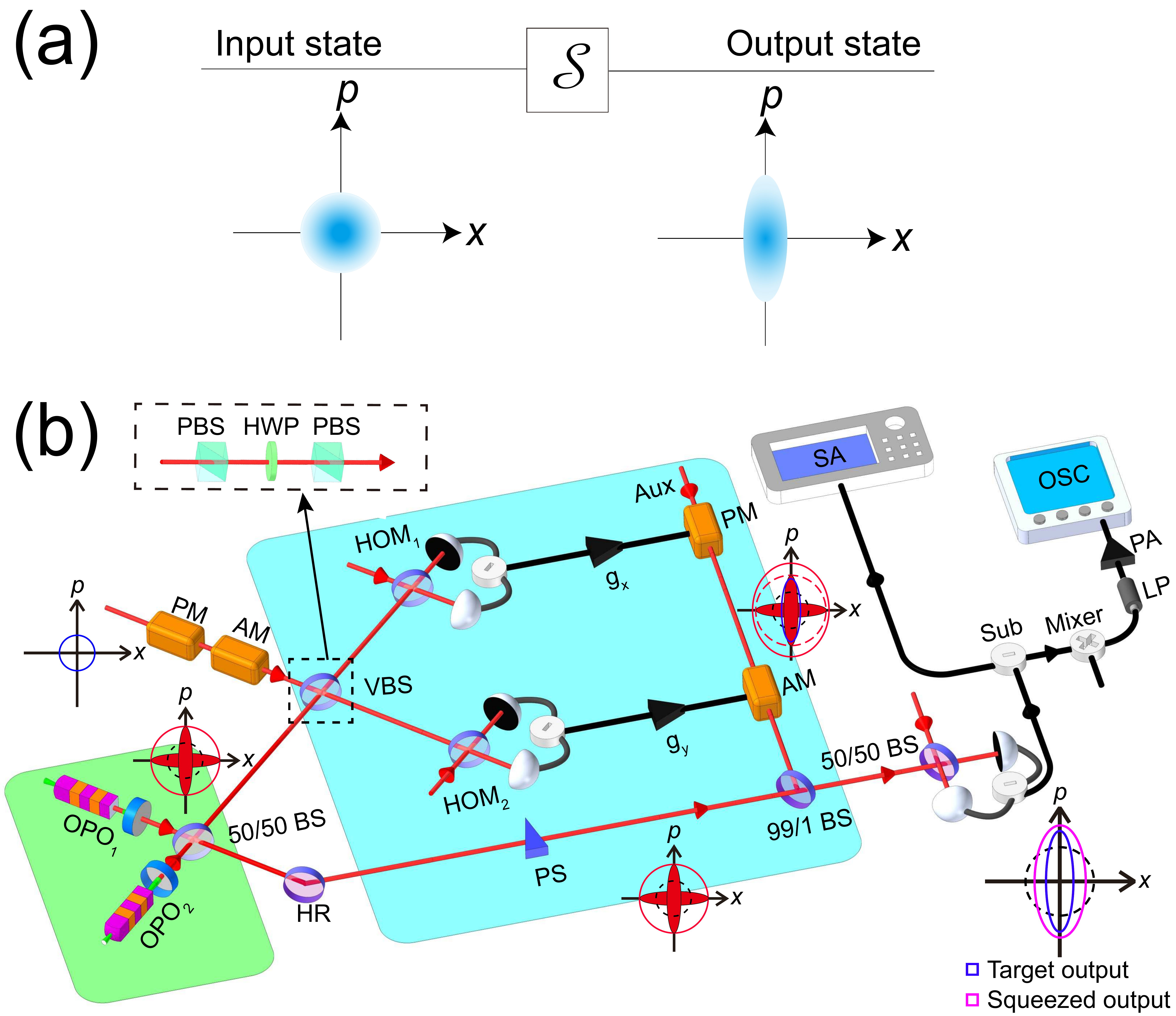}
  \caption{Schematic illustration of the deterministic squeezing gate. (a) Mechanism of squeezing gate $S$. (b) Experimental setup of off-line scheme with an entangled state. OPO: optical parametric oscillator; BS: beam splitter; HWP: half wave plate; PBS: polarization beam splitter; VBS: variable beam splitter; HR: high reflectivity mirror; AM: amplitude modulator; PM: phase modulator; PS: phase shifter; HOM: homodyne detection; Aux: auxiliary beam; Sub: substractor; SA: spectrum analyzer; LP: low-pass filter; PA: pre-amplifier; OSC: oscilloscope. }
  \label{Fig:setup}
\end{figure}
The principle of operation is shown in Figure \ref{Fig:setup}. We use the configuration with EPR entanglement assistance to realize the desired high-fidelity squeezing operation, which can be outlined as follows. Generally, quantum states of light can be described by the electromagnetic field annihilation operator $\hat{a}$. The associated amplitude and phase quadrature are written as $\hat{X}=(\hat{a}+\hat{a}^{\dagger})/\sqrt{2}$ and $\hat{P}=(\hat{a}-\hat{a}^{\dagger})/i\sqrt{2}$ respectively with the canonical commutator $[\hat{X},\hat{P}]=i$ ($\hbar=1$). In this teleportation-like protocol \cite{FurusawaScience,RevTele,HuoScience,localT}, input state denoted as $\hat{a}_{in}$ is combined with one half of the EPR beam $\hat{a}_{EPR_1}$ at a beam splitter with a variable reflectivity $R$, where the relative phase is zero. And then the amplitude $\hat{X}$ and phase $\hat{P}$ quadratures are measured by two homodyne detectors (HOMs), respectively. The extracted information from homodyne detection ($\hat{X}$ in HOM$_1$ and $\hat{P}$ in HOM$_2$) is encoded in an auxiliary beam via two independent amplitude and phase modulators with the adjustable scaling factors $g_{X}$ and $g_{Y}$, respectively (see Supporting Information \cite{SI} for more details). The other half of the EPR beam $\hat{a}_{EPR_2}$ is displaced by such an auxiliary beam by means of the 99/1 optical interference ($R_D=0.99$) with zero relative phase. The information from output state is divided into two parts, the half enters the spectrum analyzer to measure the frequency domain data, and the other is used to collect the data in time domain.

Since a EPR entangled beam used here is identified by satisfying the condition $\left \langle \Delta(\hat{X}_{EPR_1}+\hat{X}_{EPR_2})^2 \right \rangle=\left \langle \Delta(\hat{P}_{EPR_1}-\hat{P}_{EPR_2})^2 \right \rangle=e^{-2r}$, the squeezing transformation of both amplitude (Eq. \ref{amplitude_sq} for $R<1/2$) and phase (Eq. \ref{amplitude_sq} for $R>1/2$) quadratures can be realized by manipulating the variable reflectivity $R$ and optimizing the values of $g_{X}$ and $g_{Y}$, respectively, given as (see Supporting Information \cite{SI} for more details)

\begin{equation}\label{amplitude_sq}
\begin{aligned}
\left(\begin{array}{*{20}{c}}
\hat{X}_{out}\\
\hat{P}_{out}
\end{array}\right)&= \left( \begin{array}{*{20}{c}}
\sqrt{\frac{R}{1-R}}&0\\
0&\sqrt{\frac{1-R}{R}}
\end{array} \right)\left( \begin{array}{*{20}{c}}
\hat{X}_{in}\\
\hat{P}_{in}
\end{array} \right)\\
&+\left(\begin{array}{*{20}{c}}
\hat{X}_{EPR_1}+\hat{X}_{EPR_2}\\
-\hat{P}_{EPR_1}+\hat{P}_{EPR_2}
\end{array}\right).
\end{aligned}
\end{equation}

More explicitly, in contrast with the standard quantum teleportation \cite{J. Zhang, Furusawa_Tele, Furusawa, T. C. Zhang,Furusawa_telePR}, the value of $R$ is varied to manipulate the level of target squeezing, rather than being fixed at 1/2.  It is much easier to be implemented than the conventional schemes \cite{SXL_EPR,Furusawa_cluster} where the tunability is promised by the relative phase to be locked. Ideally, unit fidelity can be achieved at arbitrary target squeezing level by exploiting the EPR source with infinite degree of entanglement.

As stated above, a prerequisite for such an universal and unconditional squeezing gate is the generation of high quality optical EPR entangled state. Here, 12 dB-entangled EPR beams are experimentally produced by combining two independent squeezed beams at a 50/50 beam splitter, with the relative phase $\pi/2$ actively servo-controlled (see Supporting Information \cite{SI} for more details). 13.8 dB squeezed vacuum states of light are produced by OPOs \cite{13.8COL}, where the semi-monolithic single-resonant standing wave cavity is formed by a piezo-actuated concave mirror and the back surface of a periodically poled KTiOPO$_{4}$ (PPKTP) crystal with a dimension of 1mm$\times $2mm$\times $10mm \cite{ZYH}. In addition, to verify the universality of the squeezing gate reported here, one has to create the input states located at arbitrary location of the phase space, and this is realized by modulating the fundamental beam at 3 MHz sideband frequency using an electro-optical phase and amplitude modulators. Notice that all the relative phases, including those in the beam interference and homodyne detection, are actively stabilized during the measurement cycle in order to avoid any pollution on the non-classical source.

\section{Experimental results}

\begin{figure}
  \includegraphics[width=\linewidth]{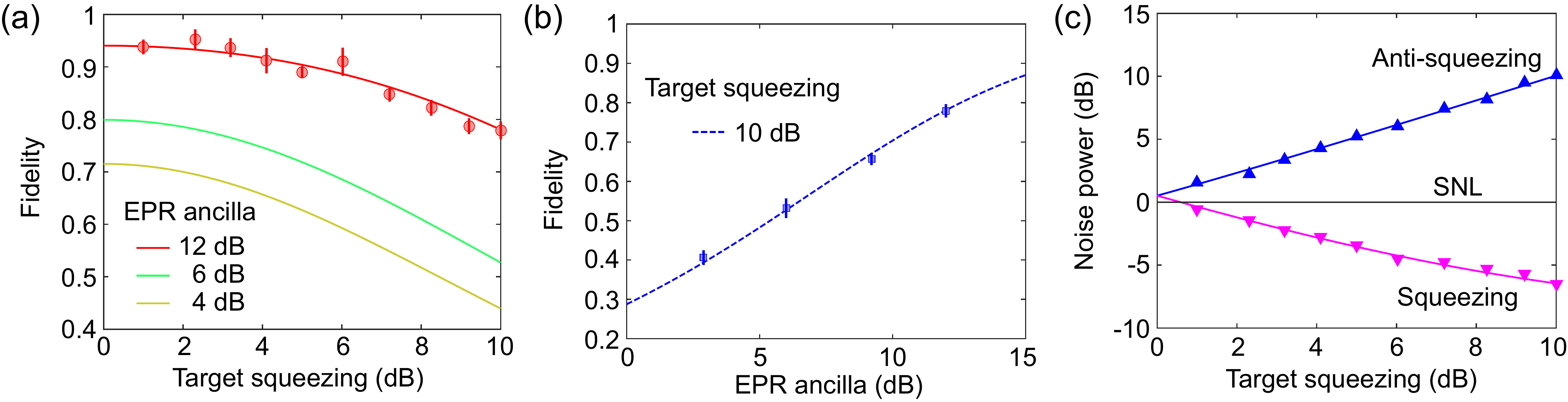}
  \caption{Fidelities as a function of the level of target squeezing (a) and EPR ancilla (b), respectively. (a) Comparison of the fidelity with different target squeezing. Red line and data points represent the theoretical and experimental results with EPR ancilla of 12 dB (this work), the yellow and green lines give the numerical results with the auxiliary resource of 4 dB and 6 dB. The error bars stand for 1 s.d. (b) Dependence of fidelity on EPR ancilla with the fixed target squeezing of 10 dB. Blue dashed line and data points show the theoretical and experimental results. (c) Squeezing and anti-squeezing level of output state as a function of target squeezing level. Experimental data (solid triangular points); theoretical curves (solid lines).}
  \label{Fig2}
\end{figure}

\begin{figure}
  \centering
  \includegraphics[width=0.4\linewidth]{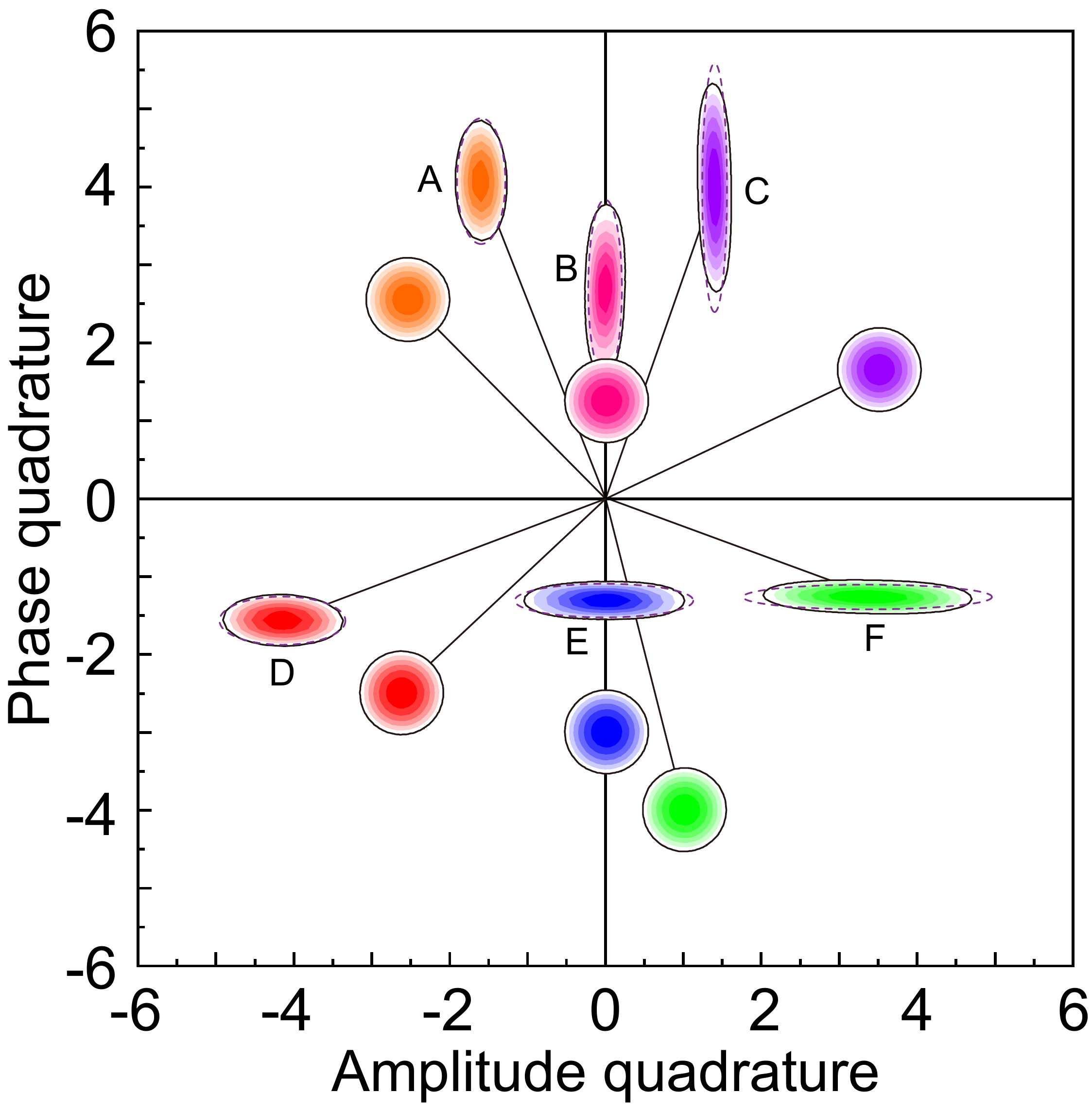}
  \caption{Phase-space diagram for the operation of squeezing gate on six different input states (A-F), located at different regions of phase space. The dashed and line ellipse represent   the theoretical predictions and  measured results of Wigner functions, respectively. The target squeezing are 4.1 dB for A and D, 7.2 dB for B and E, and 10.0 dB for C and F, respectively. A, B and C are amplitude squeezed states with the variable reflectivity $R<1/2$, while D, E and F are phase squeezed states with $R>1/2$.}
  \label{Fig3}
\end{figure}

\begin{figure}
  \centering
  \includegraphics[width=0.8\linewidth]{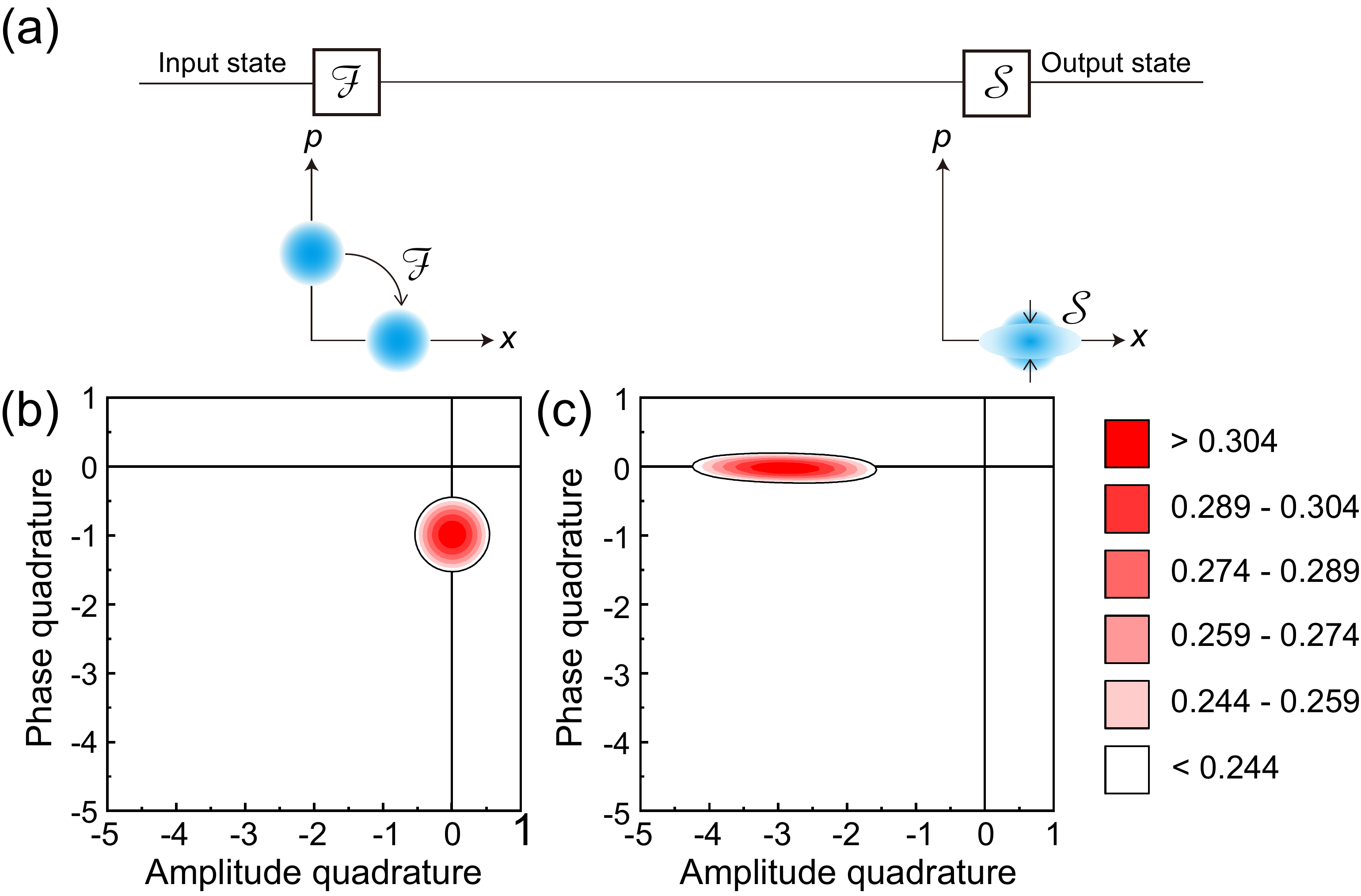}
  \caption{A complex operation with a Fourier transformation followed by a phase squeezing gate. The level of the target squeezing is 10.0 dB. (a) Schematic of the mechanism of Fourier transformation $F$ and squeezing gate $S$. (b) and (c) present the reconstructed Wigner function distribution from data collected in oscilloscope. As shown in Figure \ref{Fig4}b, the input is a $\hat{P}$-coherent state with phase quadrature 10.0 $\pm$ 0.20 dB above SNL and amplitude quadrature in SNL due to phase modulation in 3MHz. Following the Fourier transformation, amplitude and phase quadrature exchange the information with each other. It means the transformed state with amplitude variance $10.0\pm0.20$ dB above the SNL and phase variance in SNL. Finally, the squeezing in p-quadrature is implemented with target 10 dB squeezing, accompanying with $10.2\pm0.27$ dB anti-squeezing in x-quadrature. As a consequence, amplitude variance ideally approaches to the SNL while phase variance is anti-squeezed to 10 dB above the SNL, which is the interpretation of Figure \ref{Fig4}c.}
  \label{Fig4}
\end{figure}  

We proceed to evaluate the performance of our scheme. Typically, it is characterized by the fidelity, determined by the overlap between the ideal state $\left |  \psi _{i}\right \rangle$ and output state $\left |  \psi _{out}\right \rangle$, i.e. $F=\left \langle \psi _{i} \left | \hat{\rho}_{out} \right | \psi _{i}\right \rangle$ with density matrix $\hat{\rho}_{out}=\left |  \psi _{out}\right \rangle \left \langle \psi _{out} \right | $ \cite{FurusawaScience}. Experimentally, the output states are evaluated via the measurement of the noise variance of the quadrature components and also the reconstruction of Wigner functions distribution of the quantum state in phase space \cite{Lvovsky,Wigner}. Figure \ref{Fig2}a illustrates the fidelity as a function of the level of target squeezing for three values of EPR ancilla with vacuum input. Compared with previous work employing 4 dB \cite{SXL_EPR} and 6 dB \cite{PKL} EPR ancilla (yellow and green curves, respectively), our work (red points and curve) shows a big advantage, exhibiting the optimal fidelity better than 95\%. As the target squeezing is increased, the fidelity has a slower decrease. Even when the target squeezing is set up to 10 dB, the measured fidelity is still above 78\%, much better than previous work \cite{SXL_EPR,Furusawa_cluster}. Figure \ref{Fig2}b shows the fidelity as a function of the squeezing level of EPR source when fixing the target squeezing at 10 dB, indicating the significant improvement attributed to the better initial EPR source. At the high target squeezing, the high quality EPR ancilla is a key building block of quantum squeezing gate. In addition, the experimental observations are in good agreement with the theoretical predictions (red solid line in Figure \ref{Fig2}a, blue dashed line in Figure \ref{Fig2}b), demonstrating the feasibility of the presented scheme. Figure \ref{Fig2}c shows the (anti-)squeezing  level of output state as a function of target squeezing level.

As the hallmark of the unitary squeezing gate, the universality and robustness should be held, irrespective of the original location in phase space. Figure \ref{Fig3} shows the reconstructed Wigner function of six input states (marked with A-F) at different location in phase space after processing the data collected by the oscilloscope, where amplitude squeezing is operated in A-C and phase squeezing is realized in D-F. Here, the squeezing axis is set by adjusting the value of $R$, larger or smaller than 1/2, as stated above. The target squeezing for all states are 4.1 dB, 7.2 dB and 10.0 dB, respectively. We find fidelity of $0.912 \pm 0.024$ for 4.1 dB squeezing (A in orange and D in red), $0.848 \pm 0.014 $ for 7.2 dB squeezing (B in pink and E in blue), and $0.780 \pm 0.016$ for 10.0 dB squeezing ( C in purple and F in green). 

To confirm the versatility of the presented configuration, we further explore the complex operation with Fourier transformation followed by a phase squeezing gate with 10.0 dB-target squeezing as indicated in Figure \ref{Fig4}. Fourier operation is a special case ($\theta=\pi/2$) of rotation operation $R(\theta)=e^{i\theta(\hat{X}^2+\hat{P}^2)/2}$, describing the state rotated counterclockwise in phase space with an angle $\theta$. Such a complex operation is experimentally realized by exchanging the information encoded on amplitude and phase modulators, i.e. feeding forward the measurement outcome of HOM$_1$ to PM, and that of HOM$_2$ to AM. As shown in Figure \ref{Fig4}b, the input is a $\hat{P}$-coherent state with phase quadrature $10.0\pm0.20$ dB above the shot noise limit (SNL). The output state is presented in Figure \ref{Fig4}c with amplitude variance $10.2\pm0.27$ dB higher than SNL and phase variance $0.5\pm0.17$ dB above SNL, corresponding to 9.5 dB squeezing in phase quadrature. We therefore achieve a fidelity of $\mathscr{F}=0.779\pm0.017$. In this regard, by combining with a rotation operation, we can implement a universal high-fidelity squeezing gate no matter where the location of input state in phase space (see Supporting Information \cite{SI} for more details).

\section{Conclusion}
Here we report the realization of a deterministic and universal quantum squeezing gate with high-fidelity. Intriguingly, by utilizing the high quality EPR source, high fidelity can always be enabled even when the level of target squeezing increased up to 10 dB, where direct maximum measurement of 6.5 dB squeezed state is obtained. By manipulating the optical splitting ratio, the associated target squeezing is achieved, relaxing the technical challenge of relative phase locking. Its non-local property ensures the potential application in distributed quantum processor, which aims to realize the large scale quantum computer. Moreover, a complex gate sequence is operated with a high fidelity, demonstrating the feasibility of implement multi-step quantum operation using one processing unit. This work offers the opportunities to explore the hybrid discrete- and continuous-variable quantum information \cite{Ulrik_Furusawa,AndersPRA,JeongPRA,WangOE, Loock}, revealing the potential for the generation of unconventional non-classical states and even the universal quantum computation. For instance, the efficiency of Bell measurement can be boosted by inserting a CV squeezing operation before discrete variable photon counting measurement \cite{Bell,W. P. Grice Bell}. Further, by applying an extra probability filter  \cite{NLA,NLA2}, our scheme expects to achieve higher fidelity with high success probability \cite{SI}.

\medskip
\textbf{Supporting Information} \par 
Supporting Information is available from the Wiley Online Library or from the author.

\medskip
\textbf{Acknowledgements} \par 
Xiaocong Sun and Yajun Wang contributed equally. This work was supported by the National Natural Science Foundation of China (NSFC) (Grants No. 62027821, No. 11654002, No. 11874250, No. 11804207 No. 11804206 and No. 62035015); National Key R\&D Program of China (Grant No. 2020YFC2200402); Program for Sanjin Scholar of Shanxi Province;  Fund for Shanxi 1331 Project Key Subjects Construction.

\end{document}